\documentclass[aps,preprintnumbers,nofootinbib,floatfix,fleqn]{revtex4}
\usepackage{enumitem}  
\usepackage{amssymb,bm,graphicx}
\usepackage{amsmath,bbold,ulem}
\usepackage{dsfont}
\usepackage{slashed}
\usepackage[dvipsnames]{xcolor}
\usepackage[utf8]{inputenc}
\usepackage{wasysym}
\usepackage{caption}
\usepackage{subcaption}

\allowdisplaybreaks

\begin{document}

\title{Analytic Evolution of DGLAP Equations}

\author{M.~ Markovych $^1$,
A. Tandogan $^{1,2}$}
\affiliation{
  $^1$
  Department of Physics, University of Connecticut, 
  Storrs, CT 06269, U.S.A.\\
 $^2$ Department of Physics, University of Connecticut, Hartford, CT 06103}

\begin{abstract}
We present an analytical method to solve the leading order (LO) Dokshitzer-Gribov-Lipatov-Altarelli-Parisi (DGLAP) evolution equations, which describe how parton distribution functions (PDFs) vary through different energy scales.  Our approach utilizes the analytical technique that was previously employed to address the evolution of singular distribution amplitudes. The method is straightforward, mathematically transparent, and requires very little computational power. The approach involves assuming that the PDF can be expanded into a series of terms, which follow a recursion relation that we derive. To demonstrate the efficacy of our method, we utilize a toy model of PDF at initial scale. We initiate with a reasonable approximation of the experimentally calculated PDF and demonstrate that our approach yields the asymptotic behavior of the PDF.
\end{abstract}
\maketitle



\section{Introduction}
\label{Sec-1:intro}
The Deep Inelastic Scattering (DIS) experiments are a powerful tool for studying the structure of protons and neutrons. In a typical DIS experiment, high-energy electrons or muons are scattered off protons or neutrons, and the scattered particles are detected and analyzed. The DIS experiments revealed that protons and neutrons are made up of quarks and gluons, which are collectively known as partons. The partons carry a fraction of the proton or neutron's momentum, and their distributions inside the proton or neutron can be described by parton distribution functions (PDFs).

The evolution of PDFs with respect to the energy scale of the scattering process is described by the Dokshitzer-Gribov-Lipatov-Altarelli-Parisi (DGLAP)  equations \cite{gribov1972deep, dokshitzer197746, altarelli1977asymptotic}. These equations were independently derived by four groups of physicists in the 1970s and 1980s, and they are a cornerstone of perturbative quantum chromodynamics (QCD).

The DGLAP equations describe how the PDFs change as the energy scale of the scattering process changes. They are based on the idea that partons can emit and absorb gluons, which can change their momentum and spin. The DGLAP equations take into account the probability of these emissions and absorptions, and they provide a systematic way to calculate the evolution of the PDFs. The DGLAP equations are important for understanding a wide range of high-energy processes, including DIS, hadron collisions, and deep-inelastic scattering of heavy ions. They are also crucial for precision measurements at particle accelerators such as the Large Hadron Collider (LHC). A precise knowledge of PDFs plays also an important role for nuclear and particle physics experiments at Jefferson Lab, BNL, Fermi Lab and the future Electron-Ion Collider.

There exist some analytical methods to solve the DGLAP evolution equations and there are advanced and widely used numerical methods which extract the PDFs from the experimental data, e.g. CTEQ \cite{Hou_2021, Dulat_2016}, MSTW 2008 \cite{Martin_2009}, MRST2001 \cite{Martin_2002},  GRV98 \cite{Jimenez_Delgado_2009, GL_CK_2007, Gl_ck_1998} up to higher-order corrections. Mellin transformation \cite{Gluck:1989ze, Blumlein:1998if}, Laguerre method \cite{Furmanski:1981ja, Toldra:2001yz}, and ``brute-force'' iterations \cite{Cabibbo:1978ez} are some of the analytical methods to solve the DGLAP equation. There are various other analytic calculations in the literature \cite{Ball:1994kc, Kotikov:1998qt, Mankiewicz:1996sd}. The solutions to these equations provide a theoretical prediction for the PDFs that can be compared with experimental data to test our understanding of QCD and extract information about the structure of the proton including extractions of $\alpha_s$ \cite{Workman:2022ynf}.

In this paper, the aim is to provide a brief overview of an analytical method for the solution of the DGLAP equations, including the theoretical framework and the mathematical techniques involved in their solution. The method presented was previously applied to the evolution of singular distribution amplitudes (DAs) and singular generalized distribution amplitudes in Ref.\cite {Radyushkin:2014rka}. The analytical method was successful in describing the evolution of singular DAs where the standard approach of Gegenbauer polynomials is insufficient. The approach is uncomplicated and easy to follow mathematically. The PDFs are expressed by a Taylor series expansion in $t$ where $t= 2 C_F\, \ln \ln (\mu/\Lambda_{QCD})/b_0$ is the leading logarithm QCD evolution parameter.

The paper is organized in the following manner: in section \ref{basics}, we present the basics of the DGLAP evolution equations. Sections \ref{sec:sing} and \ref{sec:nonsing} are dedicated to the analytical method and its application to a toy model. In section \ref{sec:conc}, we conclude the paper with our findings and present an outlook for future studies. 

\section{\normalsize \bf Basics of Evolution Equations} \label{basics}

DGLAP evolution equations are a cornerstone of perturbative QCD calculations for many high-energy hadronic processes. These equations describe the evolution of the PDFs with increasing resolution scale, or equivalently, with decreasing probing distance. The evolution is governed by the QCD splitting functions, which describe the probability of a parton splitting into two daughter partons of lower momentum fractions. The non-singlet DGLAP equation in LO has the following form,
\begin{align}
    \frac{\partial}{\partial t}f(x,t)=\int_{x}^{1} \frac{dy}{y} \left[P(y)\right]_+ f(\frac{x}{y},t)\, \,.  
    \label{Ev_eq}
\end{align}
We have rearranged the evolution equation as the following, in order to make the complete analogy to Ref.\cite{Radyushkin:2014rka},
\begin{align}
    \frac{\partial}{\partial t}f(x,t)=\int_{x}^{1} \frac{dy}{y} \left[P(x,y)\right]_+ f(y,t)\, \,.  
    \label{eq2}
\end{align}
where $ P(x,y)$ is regulated by the ``$+$''- prescription as 
 \begin{align}
    \left[P(x,y)\right]_+=P(x,y) -\delta(y-x)\int_{0}^{1}P(z,y)\,dz. 
 \end{align} 
In its explicit form, the evolution equation is
 \begin{align}
    \frac{\partial}{\partial t}f(x,t)=\int_{x}^{1} \frac{dy}{y} \left[P(x,y)f(y,t)-P(y,x)f(x,t)\right] \, \,.  
    \label{rewritten_ev}
\end{align}
It is clear that the singularities cancel each other, even though the form of the integral does not look like the ``+''-prescription with respect to the integrating variable, $y$. By adding and subtracting $P(x,y) f(x,t)$, we can write the evolution equation as the following structure in which the
 ``$+$''-prescription acts on the integrating variable, 
\begin{eqnarray}\label{main_eq}
\int_{x}^{1} P(x,y)\left[f(y,t)-f(x,t)\right]\,\frac{dy}{y} \,-\, f(x,t)\int_{x}^{1}\left(\frac{P(z,x)}{z}-\frac{P(x,z)}{x}\right)\,dz \,.
\end{eqnarray}
The second term in Eq. (\ref{main_eq}) is also finite, 
\begin{align}\label{second_int}
  A(x)=-\int_{x}^{1}\left(\frac{P_s(z,x)}{z}-\frac{P_s(x,z)}{x}\right)\,dz \,. 
\end{align}
Following the step in \cite{Radyushkin:2014rka}, we take the Ansatz
\begin{equation}\label{ansatz}
f(x,t)=e^{A(x)t}F(x,t)\,,
\end{equation}
and obtain the following equation for $F(x,t)$
\begin{equation}\label{F_ev}
\frac{\partial}{\partial t}F(x,t) = \int_{x}^{1}P(x,y)\left( e^{t\left[A(y)-A(x)\right]}(F(y,t)-F(x,t)\right)\,\frac{dy}{y}\,.
\end{equation}
This equation does not have the second term in the integrand which existed in Eq. (\ref{main_eq}).

 We proceed by representing the function $F(x,t)$ in Taylor series.
\begin{equation}\label{taylor_exp}
F(x,t)=\sum_{m=0}^{\infty}\frac{t^m}{m!}\rho_m (x)
\end{equation}
Using this new definition of $F(x,t)$, we can rewrite Eq. (\ref{F_ev}) and obtain a recursion relation that will allow us to solve for the $\rho_m (x)$ terms:
\begin{align}\label{rho}
 \sum_{n=0}^{\infty}\frac{t^n}{n!}\rho_{n+1} (x) =  \int_{x}^{1}\frac{dy}{y}P(x,y)\left[\sum_{j=0}^{\infty}\frac{t^j}{j!}\left[A(y)-A(x)\right]^j \sum_{m=0}^{\infty}\frac{t^m}{m!}\rho_m (y) - \sum_{n=0}^{\infty}\frac{t^n}{n!}\rho_n (x)\right]
\end{align}
The double sum in the integrand can be simplified if we introduce the substitution $n=j+m$.
\begin{align} \label{two_sum}
 \sum_{n=0}^{\infty}\sum_{m=0}^{n}\frac{t^n\left[A(y)-A(x)\right]^{n-m}}{m!(n-m)!}\rho_m\ (y) = \sum_{n=0}^{\infty}\frac{t^n}{n!}\sum_{m=0}^{n}\frac{n!\left[A(y)-A(x)\right]^{n-m}}{m!(n-m)!}\rho_m\ (y)
 \end{align}
The second sum runs from 0 to n because $0\leq n-m$. On the right-hand side of Eq.~(\ref{two_sum}), we separate out the term $t^n/n!$ from the sum on m, which generates an extra factor of $n!$ in the numerator and denominator. This is done with the intention of clarifying that once we substitute the double sum into Eq. (\ref{rho}) and pull out the sum over n the following recursion relation is obtained,
\begin{equation}\label{eq11}
\rho_{n+1}(x)=\int_{x}^{1}\frac{dy}{y}P(x,y)\left[\sum_{m=0}^{n}\frac{n!\,}{m!(n-m)!}\left[A(y)-A(x)\right]^{n-m}\rho_m(y)-\rho_n (x)\right]\,.
\end{equation}
The first few iterations of this recursion relation corresponding to the terms are the following,
\begin{align}
\rho_{1}(x)&=\int_{x}^{1}\frac{dy}{y}P(x,y)\left[\rho_{0}(y)-\rho_{0}(x) \right]\,,\\
\rho_{2}(x)&=\int_{x}^{1}\frac{dy}{y}P(x,y)\left[\left(A(y)-A(x)\right)\rho_{0}(y)+\rho_{1}(y)-\rho_{1}(x) \right]\,, \\
\rho_{3}(x)&=\int_{x}^{1}\frac{dy}{y}P(x,y)\left[\left(A(y)-A(x)\right)^{2}\rho_{0}(y)+2\left(A(y)-A(x)\right)\rho_{1}(y)+\rho_{2}(y)-\rho_{2}(x) \right] \,.   
\end{align}  
where $\rho_0(x)$ is the PDF at the initial scale.
 In this manuscript, the non-singlet evolution kernel in LO is considered,
\begin{equation}\label{full_kernel}
    P(x,y)= \left[\left(\frac{2x/y}{1-x/y}\right)+\left(1-\frac{x}{y}\right)\right]\theta\left(x/y<1\right)\,.
\end{equation}

In the following section, we separate the kernel into two parts,i.e., the singular and non-singular parts. We investigate the evolution equations corresponding to these parts of the kernel separately to demonstrate the evolution method in an instructive way.

\section{\normalsize Evolution with the Singular Part of the Kernel} \label{sec:sing}
The singular part of the non-singlet evolution kernel is
\begin{equation}\label{eq5}
    P_s(x,y)= \left[\left(\frac{2x/y}{1-x/y}\right)\right]\theta\left(x/y<1\right)\,.
\end{equation}

In this section, we compute  Eq.~(\ref{F_ev}) by substituting the full kernel with solely the singular part, i.e.,  $P(x,y)\rightarrow P_s(x,y)$. This means that we only solve for a part of the PDF, namely $f_s(x,t)$. Since differentiation and integration are linear operations we can sum the results from the singular and non-singular parts.\\
\\
Now, if we substitute Eq.~(17) into the expanded- second integral term in Eq.~(\ref{main_eq}), we obtain
 \begin{align}
   A_s(x)=-\int_{x}^{1}\left(\frac{P_s(z,x)}{z}-\frac{P_s(x,z)}{x}\right)\,dz
     = 2+2\ln(\bar{x})\,.
\end{align}
Here, we define $\bar{x}=1-x$ for brevity. 

Next, we have to find a suitable candidate for $\rho_0$ so that we can calculate the succeeding $\rho_n$'s where $n\geq 1$. The choice of this term will come from related physics. We chose the initial energy PDF to be $f(x,0)=F(x,0)=\rho_0=(1-x)^3$. This choice dos not effect any outcome of the method. It is chosen to demonstrate the results. One can choose any PDF at an initial scale.


The first iteration of the recursion relation yields
\begin{eqnarray}
\rho_1(x)=\int_{x}^{1}\frac{dy}{y}P_s(x,y)(\rho_0 (y)-\rho_0 (x))
=5x-8x^2+3x^3+2x(x^2-3x+3)\ln(x)\,.   
\end{eqnarray}
Expressions of $\rho_2(x)$ and $\rho_3(x)$ are only graphically presented because of the lengthiness of the analytical expressions.

\begin{figure}[h!]
\centering
    \begin{subfigure}[b]{0.45\textwidth}
        \centering
        \includegraphics[width=\textwidth]{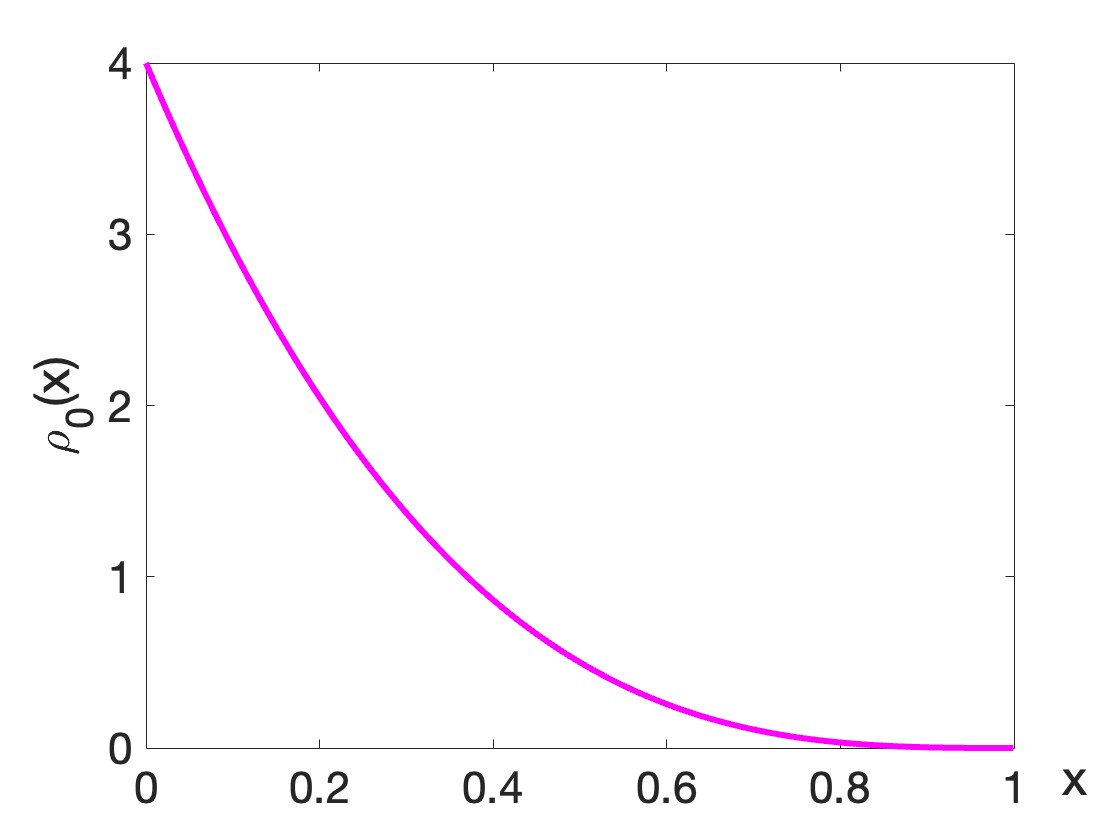}
        \caption{}
        \label{}
    \end{subfigure}
    \hspace{-0.5cm}
    \begin{subfigure}[b]{0.45\textwidth}
        \centering
        \includegraphics[width=\textwidth]{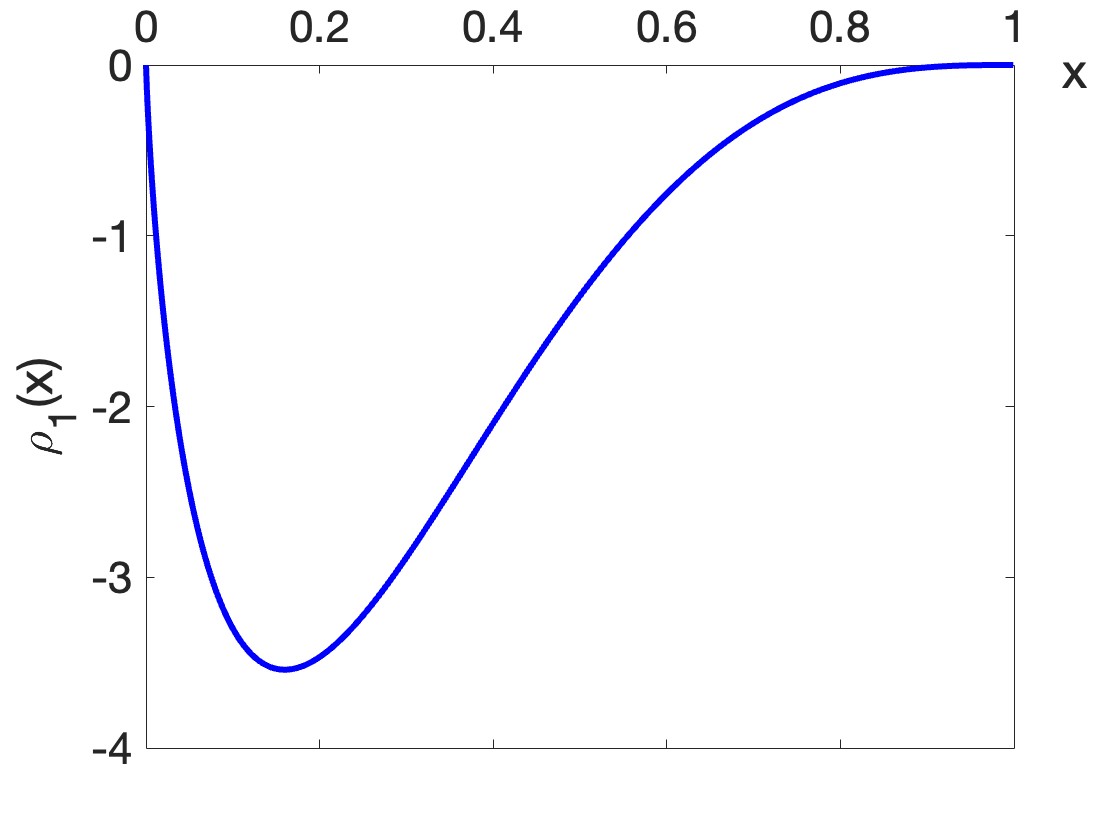}
        \caption{}
        \label{}
    \end{subfigure}
   \hspace{-0.5cm}
    \begin{subfigure}[b]{0.45\textwidth}
        \centering
        \includegraphics[width=\textwidth]{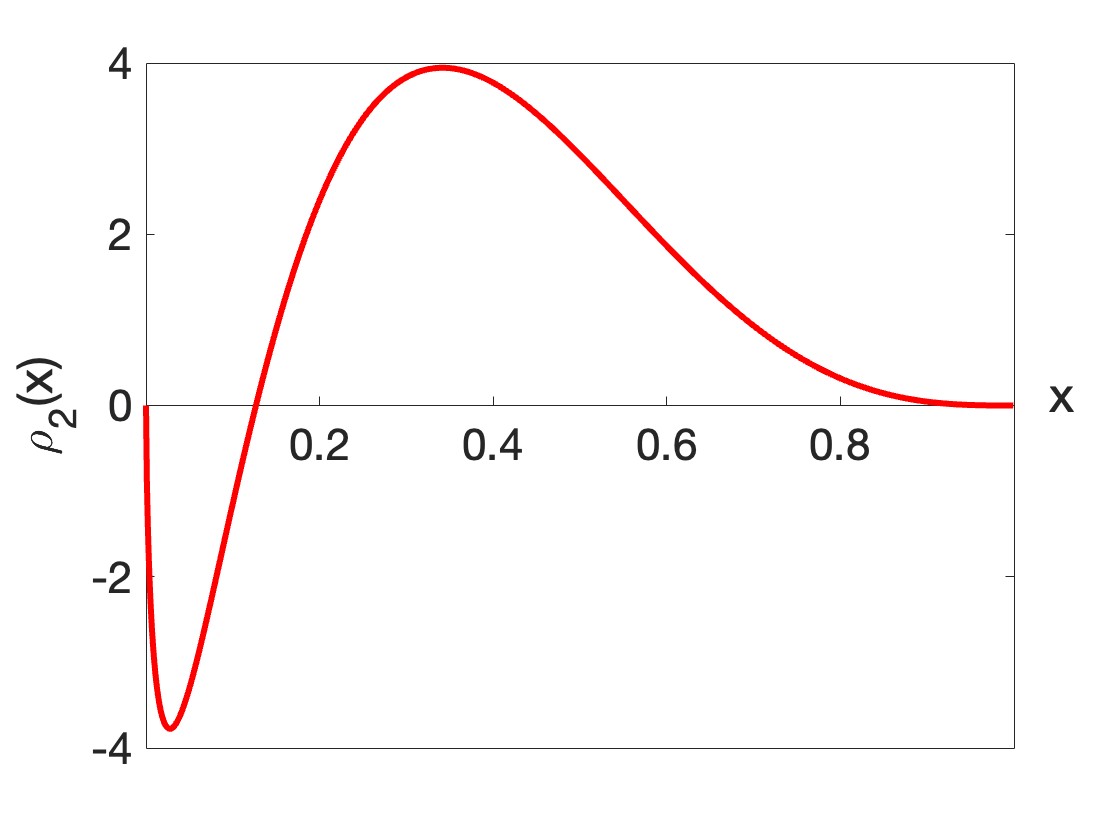}  
        \caption{}
        \label{}
    \end{subfigure}
    \hspace{-0.5cm}
    \begin{subfigure}[b]{0.45\textwidth}
        \centering
        \includegraphics[width=\textwidth]{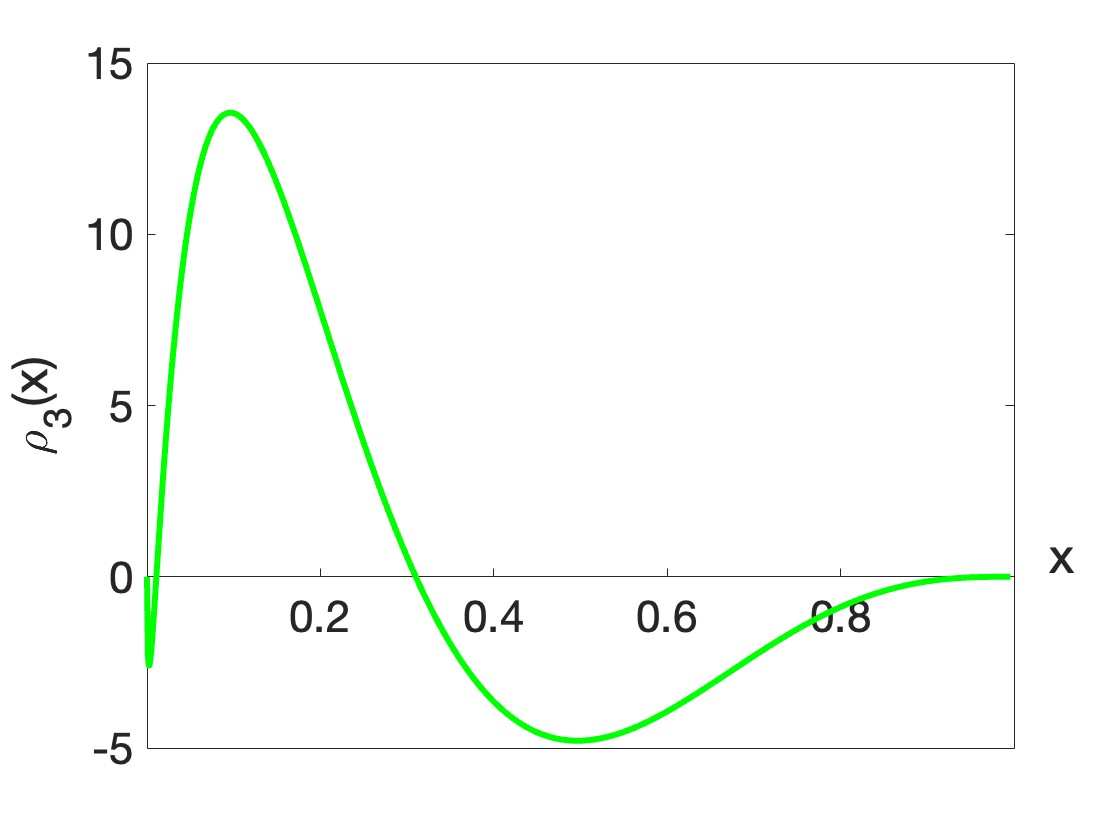} 
        \caption{}
        \label{}
    \end{subfigure}
    \caption{First three $\rho$ terms generated from singular part of the kernel}
    \label{sing_expansion}
\end{figure}

With the calculated $\rho_n$ terms, we can reconstruct the solution to the PDF using Eq.~(\ref{taylor_exp}). Inserting the iterations into Eq.~(\ref{ansatz}), we have the parton distribution function $f(x,t)$ in terms of expansion components,
\begin{equation}\label{}
f_{s}^{(3)}(x,t)=e^{2(1+\ln(\bar{x}))t}\left[\rho_0(x)+t\rho_1 (x)+\frac{t^2}{2!}\rho_2 (x) +\frac{t^3}{3!}\rho_3 (x) \right]\,.
\end{equation}
The Ansatz involves the summation of $(t \ln \bar{x})^N$ terms to an infinite order, while the series over $\rho_n(x)$ is limited to a finite order $N$. Consequently, the approximations obtained for $f^{(N)}(x,t)$ are not normalized to 1.  The normalization with three expansion components is given by
\begin{align} \label{normalization}
 I^{(3)}(t)=\int_0^1 f_s^{(3)}(x,t) dx   
\end{align}
If we do the standard normalization integral $I^{(N)}(t)$ for a consecutively increasing number of included $\rho_N$ terms the curves gradually approach 1, as expected.

\begin{figure}[h!]
    \centering
    \includegraphics[width=12cm]{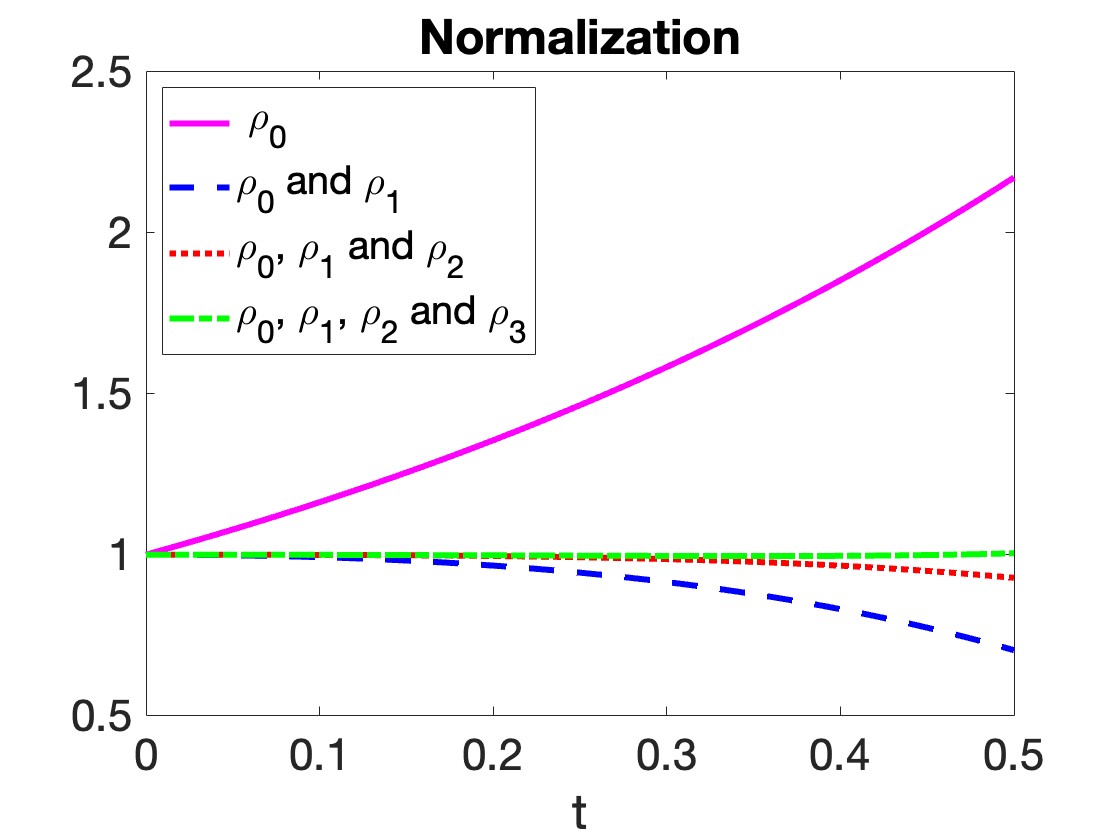}
     \caption{Plots of $\int_{0}^{1}f_{s}(x,t)dx$ including successively more $\rho$ terms as higher order corrections.}
     \label{sing_norm}
    \end{figure}
  The normalization curve in Fig. \ref{sing_norm} already approaching to zero with $N=3$ iterations included, suggesting that $I^{(3)}(t)=1+\mathcal{O}(t^4)$. This shows that the PDF, in its entirety, should satisfy the normalization condition if we simply start with the normalize $\rho_0$ term as a PDF at initial scale and then continue adding the expansion components. In other words, we expect that $\lim_{N\to\infty}I^{(3)}=1$ for all t. The normalization of the singular part up to three expansion components are depicted in Fig.~\ref{sing_norm}. The normalization integrals are computed numerically.
 
\begin{figure}[h!]
\centering
    \begin{subfigure}[b]{0.47\textwidth}
        \centering
        \includegraphics[width=\textwidth]{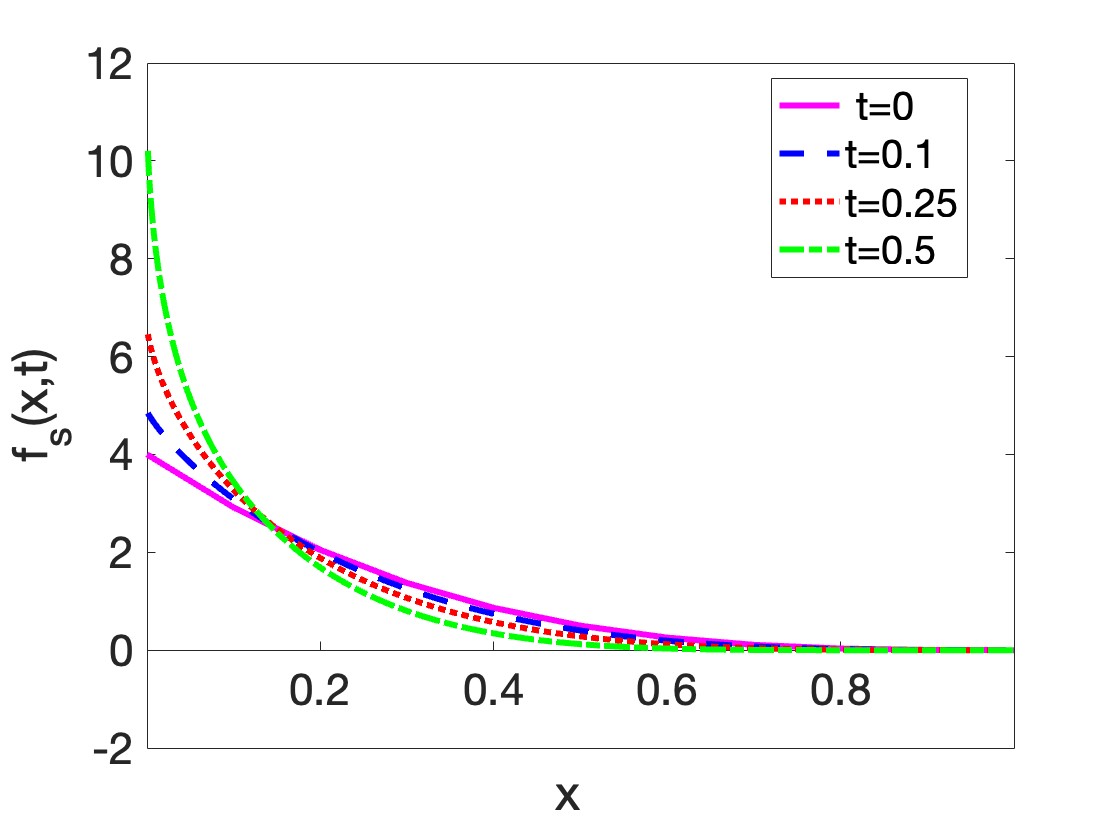}
        \caption{}
    \end{subfigure}
    \hfill
    \begin{subfigure}[b]{0.47\textwidth}
        \centering
        \includegraphics[width=\textwidth]{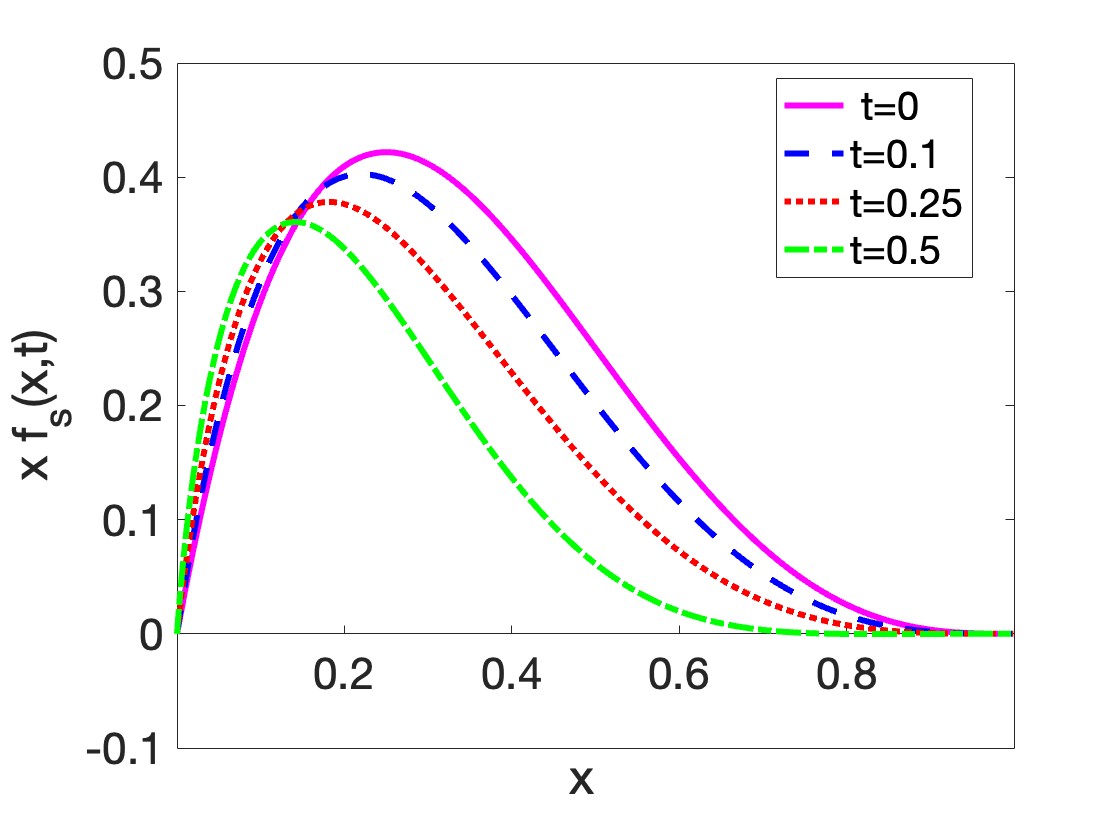}  
        \caption{}
    \end{subfigure} 
    \caption{Plots of the $f_{s}(x,t)$ and $xf_{sing}(x,t)$ at several choices of constant t, showing evolution in energy.}
    \label{sing_slices}
    \hfill
\end{figure}

 Finally, in Fig.~\ref{sing_slices}  the evolution with singular part of the non-singlet kernel is shown. The evolution of $f_s (x,t)$, in energy, meaning in the variable $t$ is presented. Despite some of the $\rho_i$ functions being negative, the evolved PDF is always positive. 
\\

\newpage
\section{Evolution with the non-singular part of the kernel}\label{sec:nonsing}
It follows from Eq.~(\ref{full_kernel}) that the non-singular part of the kernel would be

\begin{equation}\label{}
    P_{ns}(x,y)= \left[\left(1-\frac{x}{y}\right)\right]\theta\left(y-x\right)\,.
\end{equation}
\indent The same process can be repeated as before to calculate the PDF corresponding to the non-singular part. We begin by evaluating the second integral in Eq.~(\ref{second_int}) and determining the Ansatz for this part of the evaluation.
\begin{align}
A_{ns}(x)= -\int_{x}^{1}\left(\frac{P_{ns}(z,x)}{z}-\frac{P_{ns}(x,z)}{x}\right)\,dz
     = \frac{3}{2}-x-2\ln(x)
\end{align}     
The calculation of this part is fairly straightforward. By the linearity of integration, it turns out that the argument of the exponential is the direct sum of the result of this integral and the analogous one from the singular part.
\section{Evolution with full kernel}\label{sec:fullkernel}
As performed in previous sections, we now substitute the total kernel into Eq. (\ref{second_int}) and follow the same procedure in Sec.~\ref{sec:sing}. Adding the results from the singular and non-singular parts we obtain the new expression for $A(x)$ to be:
\begin{equation}\label{}
   A(x)=- \int_{x}^{1}\left(\frac{P(z,x)}{z}-\frac{P(x,z)}{x}\right)\,dz \,= \frac{1}{2}+x-\ln(x)+2\ln(\bar{x})\,.
\end{equation}
 The $n=1$ expression evaluates to the following:
\begin{eqnarray}
\rho_1(x)&&=\int_{x}^{1}\left[\frac{y-x}{y^2}+\frac{2x}{y(y-x)}\right]\left(\bar{y}^3-\bar{x}^3\right)dy \nonumber\\
&&= \frac{1}{6}(-1+x)(1+3x)(11+x(-7+2x))- x(6+(-3+x)x)\ln(x)\nonumber\\
&&+x(5-8x+3x^2+2(3+(-3+x)x)\ln(x))
\end{eqnarray}

\indent As before we do not show the results of the subsequent $\rho_i$ terms as they are very lengthy. Instead, we present them in graphical form.
We omit the $\rho_0(x)$ term as it is the same as in Fig.~\ref{sing_expansion} since it is the PDF at initial scale.
\begin{figure}[h!]
    \centering
    \begin{subfigure}[b]{0.32\textwidth}
    \includegraphics[width=\textwidth]{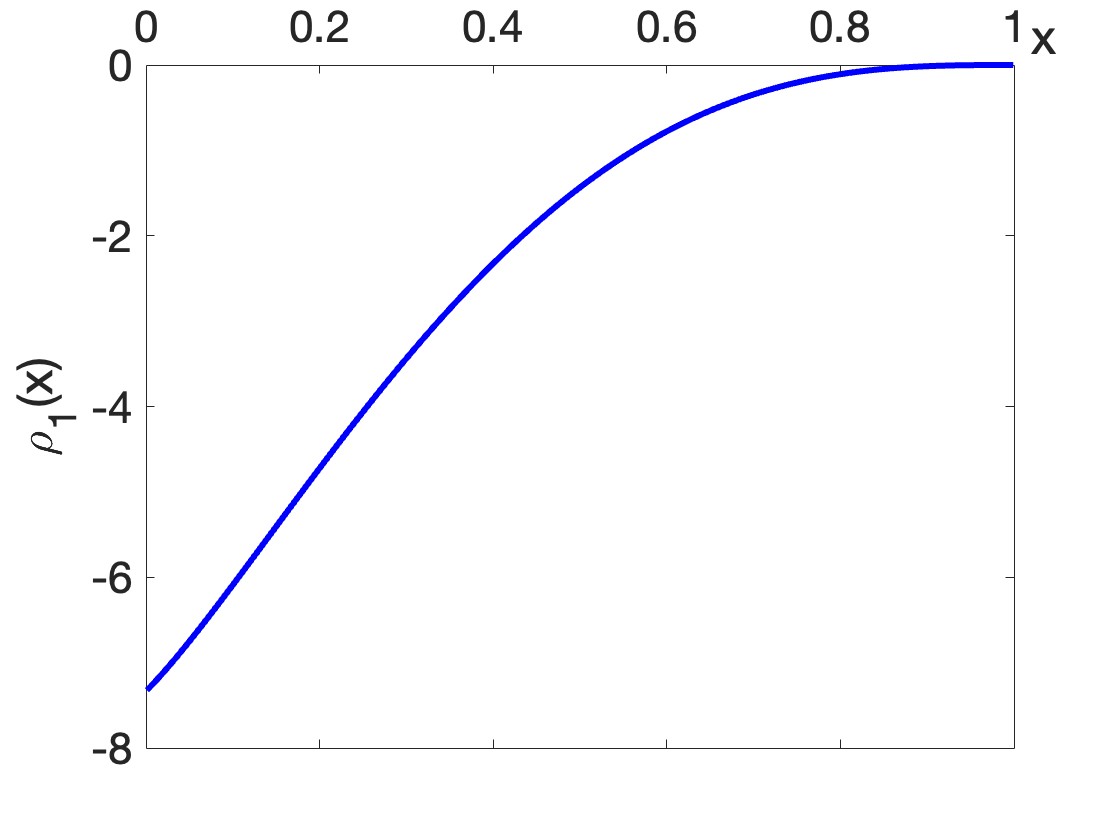}
    \caption{}
    \label{}
    \end{subfigure}
  \hspace{-0.5cm}
    \begin{subfigure}[b]{0.32\textwidth}
    \includegraphics[width=\textwidth]{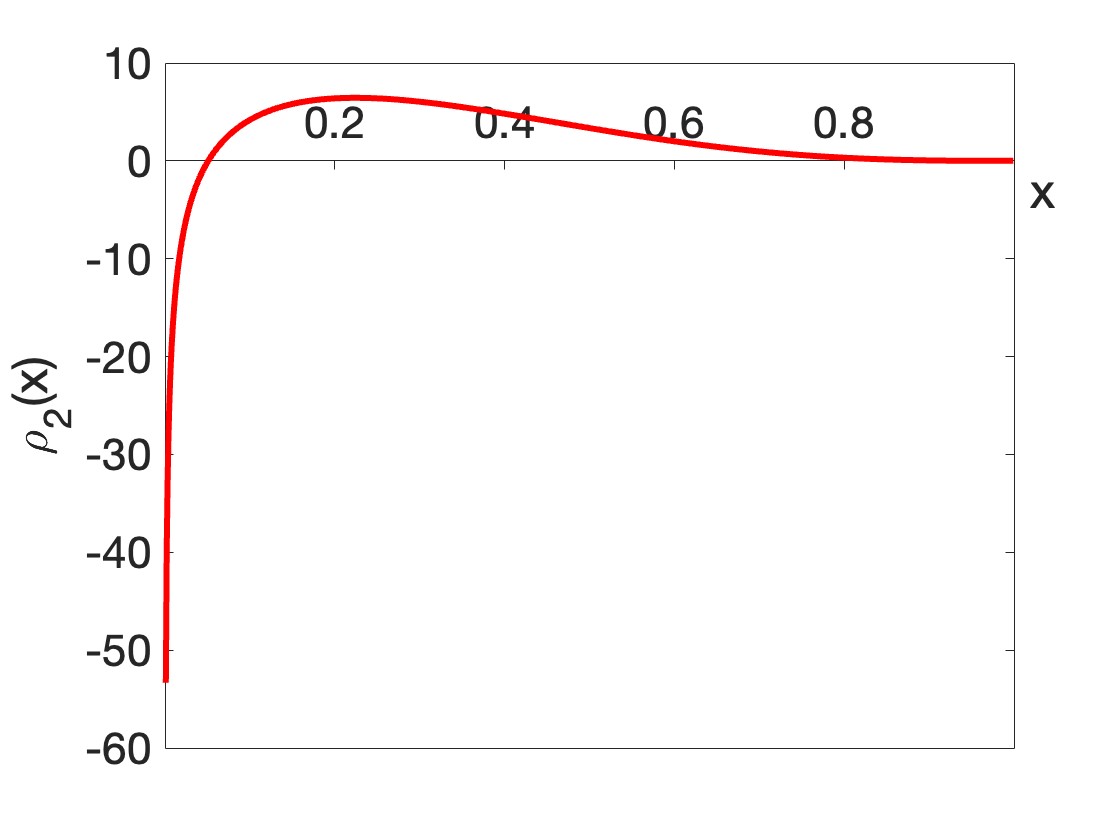}
    \caption{}
    \label{}
    \end{subfigure}
    \hspace{-0.5cm}
    \begin{subfigure}[b]{0.32\textwidth}
    \includegraphics[width=\textwidth]{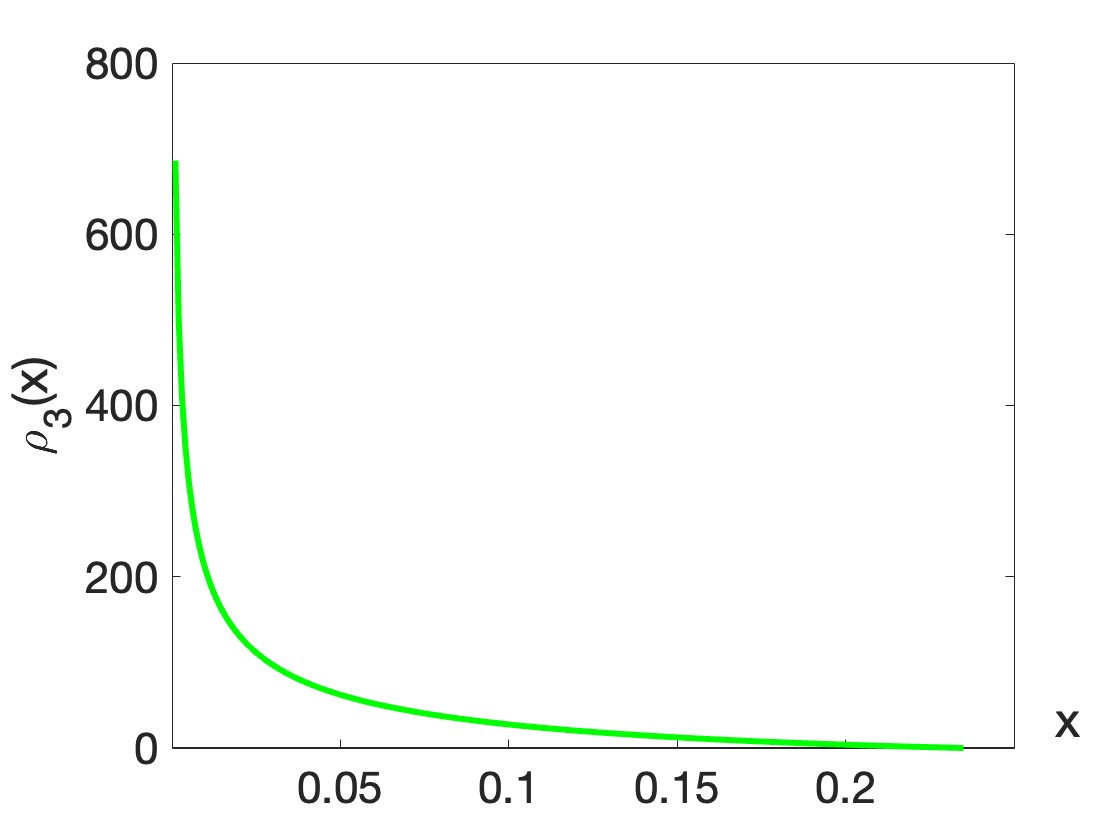}
    \caption{}
    \label{}
    \end{subfigure}
    \caption{First three iterations for total kernel}
\end{figure}

 We now move on to normalize the entirety of our solution which takes the same basic form as  Eq. (20) except with a new $A(x)$ term in the exponential and new $\rho$ terms. Adding higher order $\rho$ terms to the series for $f(x,t)$ and integrating over $x$ yields Fig. 5.

\begin{figure}[h!]
    \centering
    \includegraphics[width=12cm]{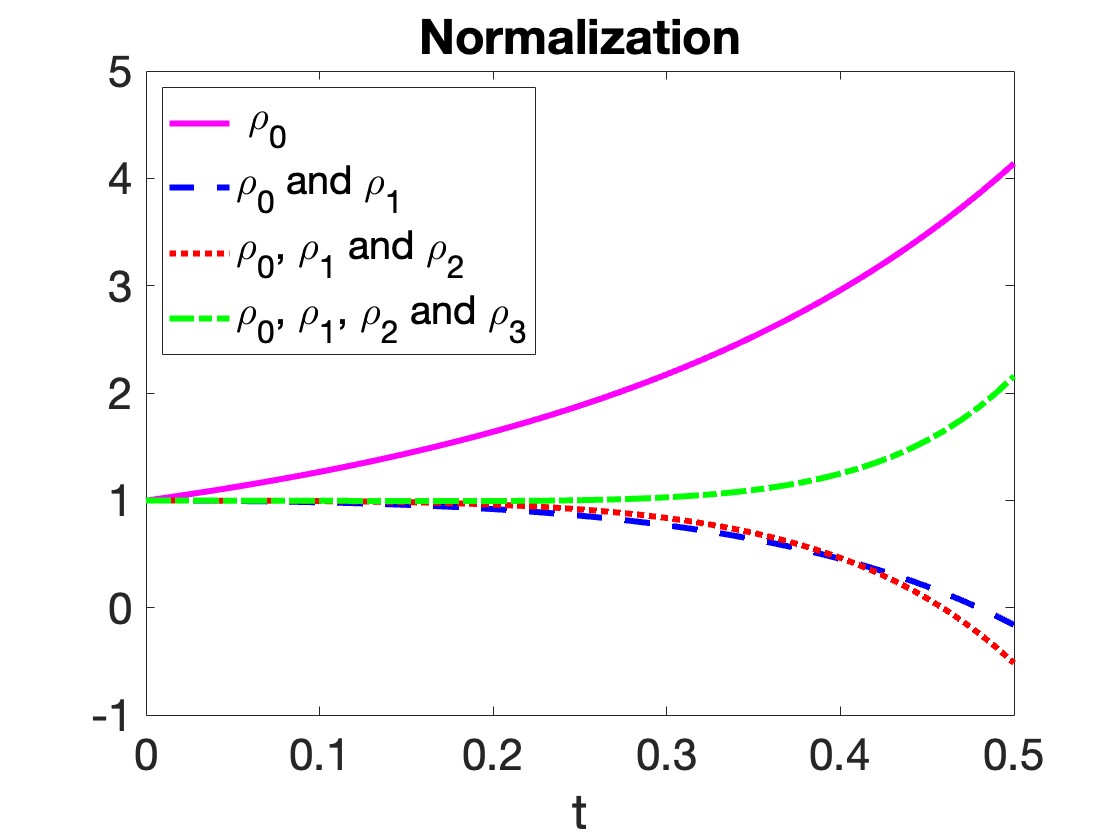}
    \caption{Comparison of the graphs of the normalized $\int_{0}^{1}f_s (x,t)dx$ including progressively higher order terms.}
    \label{tot_norm}
\end{figure}

\begin{figure}[h!]
\centering
    \begin{subfigure}[b]{0.47\textwidth}
        \centering
        \includegraphics[width=\textwidth]{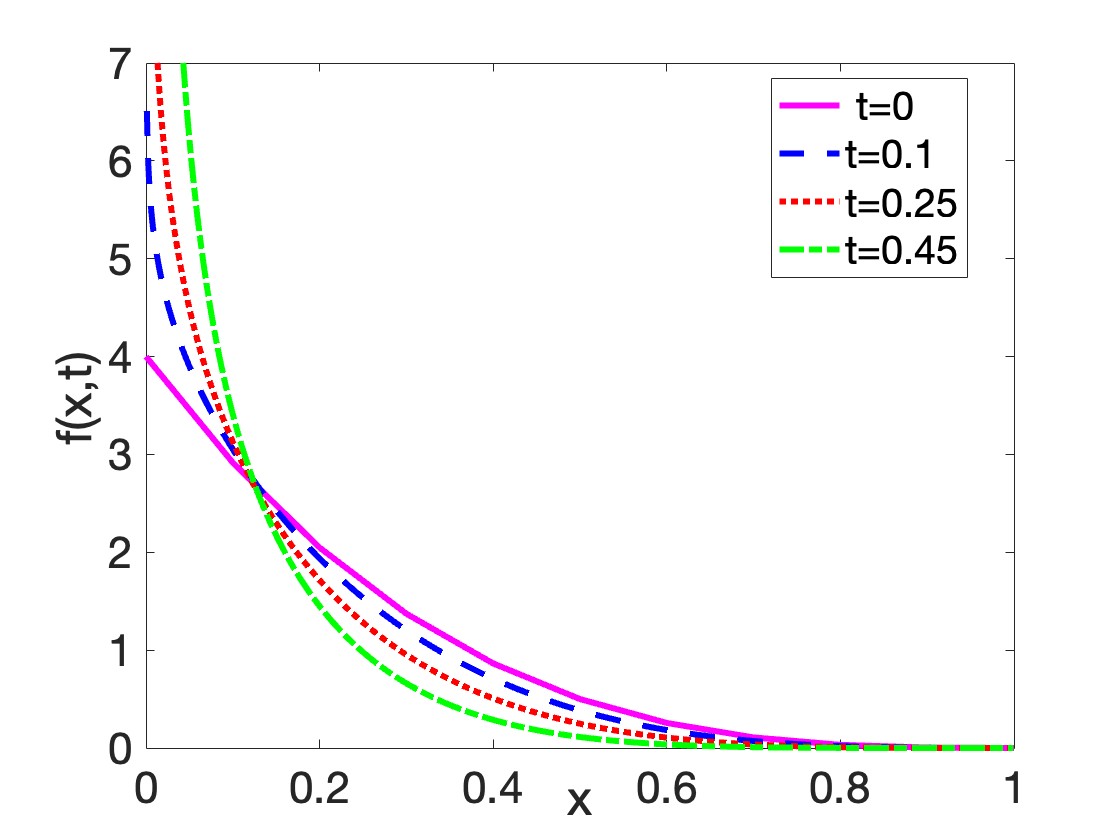}
        \caption{}
        \label{fig:ffull}
    \end{subfigure}
    \hfill
    \begin{subfigure}[b]{0.47\textwidth}
        \centering
        \includegraphics[width=\textwidth]{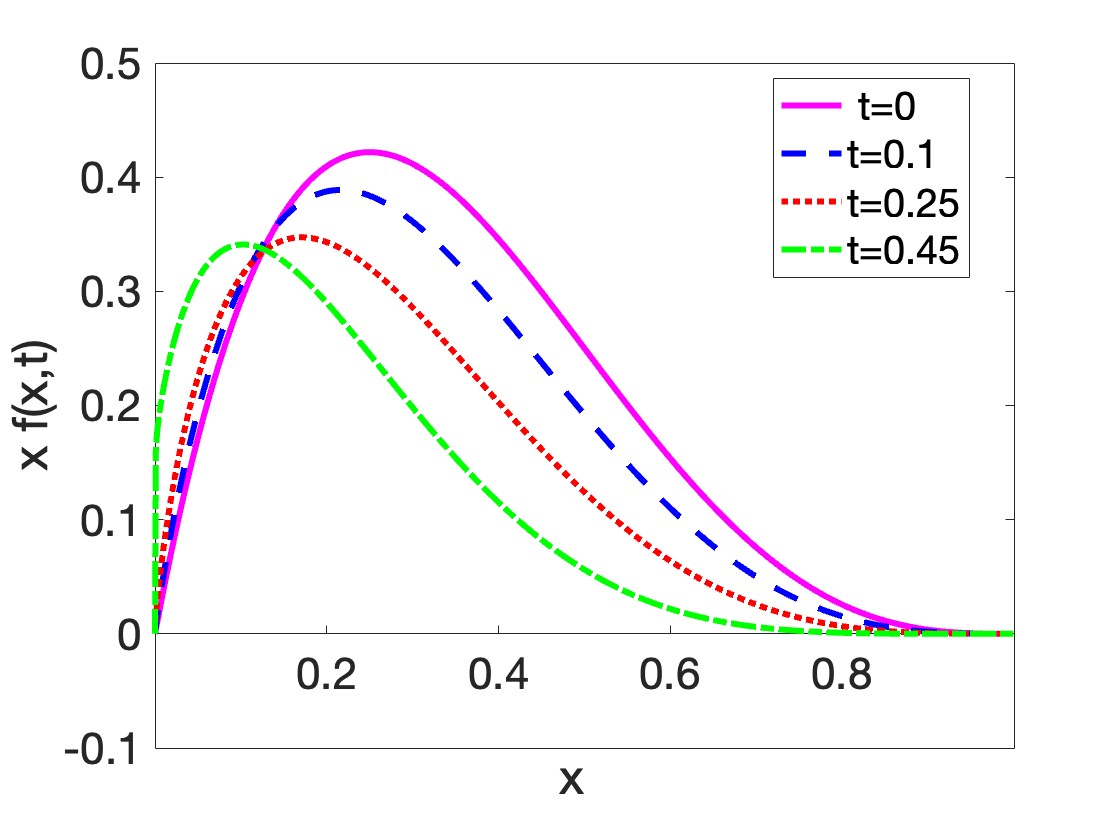}  
        \caption{}
        \label{fig:xffull}
    \end{subfigure} 
    \caption{Plots of the $f(x,t)$ and $xf(x,t)$ at several choices of constant t, showing evolution in energy.}
    \hfill
\end{figure}

\indent In Fig.~(\ref{tot_norm}), one realizes that this normalization graph doesn't get tend to approach to 1 as fast as in Fig. (2) for singular kernel. However, the trend stays the same, i.e., as more $\rho_n$ terms are added the normalization centers around 1 for a progressively longer amount of time. Although the area under the full PDF including all the calculated $\rho$ terms begins to divert from 1 closer to the value of $ t \approx 0.5$ the solutions up to $t \leq 0.4$ are reliable. The way to interpret this is that the method is effective up $ t \approx 0.5$.

In Fig.~(\ref{fig:ffull}), and~(\ref{fig:xffull}), it is noticeable that the last curve corresponding to $t=0.45$ begins to slightly veer off the path that the other curves seem to be following. This is due to the fact that the PDF at this $t$ begins to diverge around $0$. Therefore, when we multiply it by $x$ the convergence to 0 is a little less smooth. We can anticipate that for larger $t$, where the normalization steps away from 1, the PDF will not be as smooth as one would like. 
\newpage
\section{Conclusion and Outlook}\label{sec:conc}

We utilized the analytical method that was formerly used to analyze the evolution of singular distribution amplitudes and applied it to the DGLAP evolution equations. We have observed that the computation containing only the singular part of the kernel performs exceptionally well. It achieves near-perfect alignment with the constant value of 1 as shown by the normalization curve in Fig.~\ref{sing_norm}, despite having only computed terms up to 3rd order (up to $\rho_{3}(x)$). Comparing the normalization curves we see that with each additional inclusion of a higher order $\rho$ term, the correction to the previous curve gets progressively smaller. Following this pattern we can infer that if we calculated $\rho_{4}(x)$ the correction to the last normalization curve will be even smaller. This means that we indeed are successfully normalizing the PDF simply by adding more $\rho$ terms to the series. In consequence, the corresponding $f_{s}(x,t)$ and $xf_{s}(x,t)$ plots are well normalized and approach the asymptotic behavior.\\

 In the calculation with the full kernel, it is realized that although the solution's behavior isn't as ideal as for the singular portion it is nonetheless viable. The normalization curve values corresponding to the full kernel in Fig.~\ref{tot_norm} do not approach to 1 as fast as it did with only the singular part of the kernel. However, with each successive inclusion of a $\rho$ term into the normalization integral we see that result hangs tends to approach to 1 for a progressively larger amount of energy. The pattern in Fig.~\ref{tot_norm} indicates that adding higher-order terms to the series is in fact gradually normalizing the full solution properly. The increasing behavior of the final green dotted normalization curve around $t \approx 0.5$ can be attributed to the strongly increasing behavior of $\rho_{3}(x)$ around $x=0$. Due to this, the plots of $f(x,t)$ vary steeply around $x=0$ for $t \approx 0.5$. This divergent behavior around $x=0$ is expected as it is also observed in the experimentally determined PDF. However, in sight of the fact that the normalization curve isn't perfect and that we have only calculated the $\rho$ terms up to order 3, one can conclude that the evolution for higher energies $t \approx 0.5$  either requires inclusion of further expansion components, i.e., $\rho_n$'s or application of other numerical or analytical methods . It is still a viable method up to energies around $t \approx 0.5$.\\
 In summary, starting from our idealized, yet sensible, initial choice $f(x,0)=(1-x)^{3}$ we analyzed the energy-evolved PDF curves. This toy model shows that you can in practice use this method to obtain efficient and mathematically transparent calculations of energy-evolved PDFs. This alternative analytical method offers an intuitive solution to DGLAP evolution equations.

The efficiency of the analytical approach we introduced is a valuable  method for computing the evolution of Generalized Parton Distributions (GPDs), which can involve Dirac-Delta functions. In a forthcoming publication, we will describe how this method can be utilized to analyze GPD evolutions.

\section*{Acknowledgements}
We are thankful to Peter Schweitzer for discussions and support. This work was partly supported by the National Science Foundation under the Contract No. 2111490.
\bibliography{dglap_ref.bib}

\end{document}